\documentclass[prd,preprint,a4paper,superscriptaddress,nofootinbib,11pt]{revtex4}
\usepackage{amssymb}
\usepackage{amsfonts}
\usepackage{amsmath}
\usepackage{hyperref}
\usepackage[dvips]{graphicx}

\def\fs{\footnotesize}
\begin{document}

\title{\Large Accelerated Cosmological Models in \\ Modified Gravity\\
tested by distant Supernovae SNIa data.}

\date{ \today }

 \author{Andrzej BOROWIEC}
 \email{borow@ift.uni.wroc.pl}
 \affiliation{{\fs Institute of Theoretical Physics, University of Wroc{\l}aw\\
 Pl. Maksa Borna 9, 50-204  Wroc{\l}aw (Poland).}}
  \author{W{\l}odzimierz GOD{\L}OWSKI}
 \email{ godlows@oa.uj.edu.pl}
 \affiliation{{\fs  Astronomical Observatory  Jagiellonian University
\\ 30-244 Krak{\'o}w, ul. Orla 171, Poland}}
 \author{Marek SZYD{\L}OWSKI }
 \email{ uoszydlo@cyf-kr.edu.pl}
 \affiliation{{\fs  Astronomical Observatory  Jagiellonian University
\\ 30-244 Krak{\'o}w, ul. Orla 171, Poland\\
Mark Kac International Centre for Complex and Quantum System\\
Institute of Physics Jagiellonian University\\
30-244 Krakow, ul. Orla 171, Poland}}


\begin{abstract}
 Recent type Ia supernova  measurements and other astronomical
observations suggest that our universe is, at the present epoch, in
an accelerating phase of evolution. While a dark energy of
unknown form and origin was usually proposed as the most feasible
mechanism for the acceleration, there appeared some
generalizations of Einstein equations which could mimic dark energy.
In this work we investigate observational constraints on a
modified Friedmann equation obtained from the generalized Lagrangian
${\cal L}\propto R^n$  minimally coupled with matter via the Palatini
first-order formalism. We mainly concentrate on such restrictions of
model parameters which can be derived from distant supernovae and
baryon oscillation tests. We obtain confidence levels for two 
parameters ($n$, $\Omega_{m,0}$) and find, from combined analysis, that
the preferred value of $\Omega_{m,0}$ equals $0.3$.
For deeper statistical analysis  and for comparison of our model with 
predictions of  the $\Lambda$CDM concordance model one  applies Akaike and 
Bayesian information criteria of model selection. Finally, we conclude that 
the Friedmann-Robertson-Walker (FRW) model merged with a first-order 
non-linear gravity survives SNIa and baryon oscillation tests.
 \end{abstract}

\maketitle

%%%%%%%%%%%%%%%%%%%%%%%%%%%%%%%%%%%%%%%%%%%%%%%%%%%%%%%%%%%%%%%%%%%%%%%%%%%%%%%%%%%%%%%%%%%%%%%%%%%%%%%

\section{Introduction}

The recent observations of type Ia distant supernovae indicate that
our Universe is currently accelerating \cite{Riess:1998cb,Perlmutter:1998np}.
There are different proposals for explaining this phenomenon.
Some of them are based on assumptions of standard cosmological models,
which utilize  FRW metric.
Thus possible explanations include: cosmological constant $\Lambda$
\cite{Weinberg:1989,Carroll:1992}, a decaying vacuum energy density
\cite{Vishwakarma:2001}, an evolving scalar field or quintessence models
\cite{Ratra:1998}, a phantom energy (expressed in terms of the barotropic
equation of state violating the weak energy condition)
\cite{Caldwell:1999ew,Dabrowski:2003}, dark energy in the form of
Chaplygin gas \cite{Kamenshchik:2001cp}, etc.. All these conceptions
 propose some kind of new matter of unknown origin which violate
the strong energy condition. The Universe is currently accelerating
due to the presence of these dark energy components.

On the other hand, there are alternative ideas of
explanation, in which instead of dark energy some
modifications of Friedmann's equation are proposed at the very beginning.
In these approaches some effects arising from new physics like brane
cosmologies, quantum effects, anisotropy effects etc. can mimic dark
energy by a modification of Friedmann equation. Freese \& Lewis
\cite{Freese:2002sq} have shown that contributions of type $\rho^n$
to Friedmann's equation $3H^2=\rho_{eff}$,
where $\rho$ is the energy density and $n$ a constant, may describe 
such situations phenomenologically. These models (by their authors 
called the Cardassian models) give rise to acceleration, although the 
universe is flat and contains the usual matter and radiation without
any dark energy components. In the authors' opinion \cite{Freese:2002sq},
what is still lacking is a fundamental theory (like general relativity)
from which these models can be derived after postulating Robertson Walker
(R-W) symmetry. We argue that a possible candidate for such a fundamental
theory can be provided by non-linear gravity theories (for a recent review
see e.g. \cite{NO} and references therein) and, particularly, the so-called
$f(R)$- theories \cite{Cap}. It is worth  pointing out that if one imposes
the energy-momentum conservation condition then matter density is
parametrized by the scale factor (in a case of R-W symmetry), and
Cardassian term $\rho^n$ in the Friedmann equation will be
reproduced.

There are different theoretical attempts to modify gravity in order
to achieve an accelerating cosmic expansion at the present epoch.
Already in the paper by Carroll et~al. \cite{Carroll04},  one
can find interesting modifications of the Einstein-Hilbert action
with Lagrangian density ${\cal L} \propto R + f(R, P, Q)$.
Those authors have shown that in the generic case cosmological models
admit, at late time, a de Sitter solution, which is unfortunately unstable.
Moreover, Carroll et~al. have demonstrated the existence of an interesting 
set of attractors, which seem to be important in the context of the dark
energy problem.

The main goal of the present paper is to set up observational constraints on
parameters of cosmological models inspired by non-linear gravity.
The possibility of explaning cosmic acceleration in terms of
nonlinear generalization of the Einstein equation has been previously
addressed in \cite{XY,XX}.  However, these authors have not confronted 
their models with observational data. This problem has been tackled in 
\cite{yy}, where nonlinear power law lagrangians were compared with SNIa 
data and X ray gas mass fraction as well (see also \cite{mota} for a more 
general class of lagrangians). Here, we use samples of supernovae Ia
\cite{Riess:2004nr,Astier:2005} together with the baryon oscillation test
\cite{Eisenstein:2005} for stringent and deeper constraint on model
parameters. We check to which extent the predictions of our model are
consistent with the current observational data.

Severe constraints on the particular modifications of gravity
considered in this paper have been already proposed
\cite{Cap,yy,mota,Severe}. On the other hand, in the article by Clifton
and Barrow \cite{Clifton05}, strong constraints coming  from
nucleosynthesis of light elements have been  found within a
higher-order gravity. Therefore, it is possible that our model
(although a first-order), which fits SNIa data well can be ruled out
by nucleosynthesis arguments.

Assuming FRW dynamics in which dark energy is present, the basic
equation determining a cosmic evolution has the form of a generalized 
Friedmann equation
\begin{equation}
\label{eq:1}
H^{2} = \frac{\rho_{\mathrm{eff}}}{3}-\frac{k}{a^2}\ ,
\end{equation}
where $\rho_{\mathrm{eff}}(a)$ stands for the effective energy density of
several ``fluids'', parametrized by the scale factor $a$, while $k = \pm 1,0$ 
denotes the spatial curvature index. One can reformulate (\ref{eq:1}) in terms 
of density parameters $\Omega_i$ as
\begin{equation}
\label{eq:2}
\frac{H^{2}}{H_{0}^{2}} = \Omega_{\mathrm{eff}} (z) + \Omega_{k,0}(1+z)^2 \ ,
\end{equation}
where $\frac{a}{a_0}=\frac{1}{1+z}$,
$\Omega_{\mathrm{eff}}(z)=\Omega_{m,0}(1+z)^3+\Omega_{X,0}f(z)$ and
$\Omega_{m,0}$ is the density parameter for the (baryonic and dark)
matter, scaling like $a^{-3}$, while $f(z)$ describes the dark energy $X$.
For $a=a_{0}$ (the present value of the scale factor
which we further on normalize to unity), one obtains the
constraint $\Omega_{\mathrm{eff,0}} + \Omega_{k,0}=1$.

We can certainly assume that the energy density ($i=m,X$) satisfies 
the conservation condition
\begin{equation}
\label{eq:3}
\dot{\rho}_{i} = -3H(\rho_{i} + p_{i})
\end{equation}
for each component of the fluid, so that
$\rho_{\mathrm{eff}}=\Sigma\rho_i$. Then from ~(\ref{eq:2}) one gets
the constraint relation $\Sigma_i\Omega_{i,0} +\Omega_{k,0}=1$ for
the present values ($z=0$) of the density parameters.

All approaches mentioned above lead toward a description of dark energy in 
the framework of standard FRW cosmology. 
It will be demonstrated, in the next Section, that all cosmological models
of the first-order non-linear gravity which satisfy R-W symmetry, can also 
be reduced to the familiar form (\ref{eq:2}). Therefore, the effects of 
nonlinear gravity can mimic dynamical effects of dark energy.

%%%%%%%%%%%%%%%%%%%%%%%%%%%%%%%%%%%%%%%%%%%%%%%%%%%%%%%%%%%%%%%%%%%%%%%%%%%%%%%%%%%%%%%%%%%%%%%%%%%%%%%
\section{FRW cosmology and first-order non-linear gravity}
\noindent For the cosmological applications one chooses the
Friedmann-Robertson-Walker metric, which (in spherical
coordinates) takes the standard form:
\begin{equation}
g=-d t^2+a^2 (t) \Big[ {1 \over {1-k r^2}} d r^2+ r^2 \Big( d \theta^2 +\sin^2 (\theta) d \varphi^2  \Big) \Big]\ .
\label{RW1}
\end{equation}
As before, $a (t)$ denotes the scale factor and $k$ the spatial
curvature ($k=0,1,-1$). Another main ingredient of all cosmological
models is a perfect fluid stress-energy tensor, expressed by
\begin{equation}
T_{\mu \nu}=
\left(
\begin{array}{clcr}
\rho&0&0&0\\
0&\frac{pa^2 (t)}{1-k r^2}&0&0\\
0&0&pa^2 (t)r^2&0\\
0&0&0&pa^2 (t)r^2\sin^2 (\theta)
\end{array}
\right) \ .
\label{Tmunu}
\end{equation}
One requires the standard relations between the pressure $p$, 
the matter density $\rho$, the equation of state parameter $w$ 
and the expansion factor $a(t)$, namely
\begin{equation}
p=w\rho \quad , \quad
\rho=\eta a^{-3(1+w)}, \quad, \quad \eta=\mathrm{const}.
\label{pro}
\end{equation}

Let us  consider the action functional
\begin{equation}
A=A_{\mathrm{grav}}+A_{\mathrm{mat}}=\int (\sqrt{\det g}f(R)+2\kappa L_{\mathrm{mat}} (\Psi) )d^{4}x
\end{equation}
within the first order Palatini formalism \cite{XX}. In fact, from now on 
we shall assume the simplest power law gravitational Lagrangian of the form
\[
f(R)\sqrt{g} =\frac{\beta}{2-n} R^n \sqrt g \qquad (\beta \ne 0; \; n \in  {I\kern-.36em R}  ; \; n\ne 0,2) \ ,
\]
where one fixes the constant $\beta$ to be positive (it has the same dimension 
as $R^{1-n}$).  We want to point out that our model is singular for $n=2$.
As shown in \cite{XX}, such models are exactly solvable for the matter 
stress-energy tensor representing a single perfect fluid of a kind $w$
(cf. (\ref{pro})). Their confrontation with experimental data has been
performed, for a dust filled universe, in \cite{yy}. Here we attempt
to continue the analysis with newly available Astier SNIa samples and new
baryon oscillation tests. These allow us  to strengthen the admissible 
constraints on model parameters. Moreover, we extend our research to a matter
stress-energy tensor containing two components, both with $p=w\rho$: a perfect 
fluid $w= \mathrm{const}  \neq \frac{1}{3}$ and a radiation $w=\frac{1}{3}$. 
It turns out that the presence of the radiation term crucially changes the 
dynamics of our model at the early stage of its evolution. In addition, as to 
be demonstrated in Section III, although one cannot obtain better constraints 
from SNIa data, the combined analysis of SNIa and baryon oscillations offers a 
new possibility for a deeper determination of model parameters.

Following a method developed in \cite{XX}, the Hubble parameter for our model
can be calculated to be:
\begin{eqnarray}
H^2 = \frac{2n}{3(3w-1)[3w(n-1)+(n-3)]}
\left[  \frac{\kappa(1-3w) \eta_w}{\beta} \right]^{\frac{1}{n}}\,a^{-\frac{3(1+w)}{n}}+
\nonumber \\
+\frac{4n(2-n)\kappa\eta_{rad}}{3\beta[3w(n-1)+(n-3)]^2}
\left[  \frac{\kappa(1-3w) \eta_w}{\beta} \right]^{\frac{1-n}{n}}\,a^{-\frac{n+3+3w(1-n)}{n}}
- \frac{k}{a^2}\, \left[ \frac{2n}{3w(n-1)+(n-3)} \right]^2 \ .
\label{MFRn}
\end{eqnarray}
(Since radiation is already included in (\ref{MFRn}), one has to
assume $w\ne{1\over 3}$.)

It is worth pointing out that the deceleration parameter, in the 
case $k= \eta_{rad}=0$, equals to (see \cite{XX}):
\begin{equation}
q(n, w)=\frac{3(1+w)-2n}{2 n}=-1+\frac{3(1+w)}{2 n} \ .
\label{qquti}
\end{equation}
Thus, the effective equation of state parameter $w_{eff}$  is
\begin{equation}
 w_{eff}=-1+{1+w \over n} \ .
\label{weff}
\end{equation}
Let us observe that, in the case $\eta_{rad}\neq 0$, $w<\frac{1}{3}$, the same
values of $q(n, w)$ and $w_{eff}$ can also be
achieved as asymptotic values $a \mapsto \infty$.
In the early universe, when  the scale factor goes to the initial
singularity, the radiation term, in (\ref{MFRn}) scaling like
$a^{-(1+\frac{3}{n})}$, will dominate over  the dust term (scaling like 
$a^{-\frac{3}{n}}$). More precisely, if $n<0$ or $n>2$, then the negative 
radiation term cannot dominate over the matter, so that instead of the 
initial singularity we obtain a bounce. On the other hand, if $a$ goes 
to infinity, the radiation becomes negligible versus to the matter.

\noindent
In our further analysis we restrict ourselves to the case  $w=k=0$ i.e., more 
precisely, to the spatially flat universe filled with dust and radiation. Thus 
(8), (remarking once more that all $\eta$'s are positive constants) becomes
\begin{equation}
H^2=
\frac{2 n}{3(3-n)}
\left[\frac{\eta_{dust}\kappa}{\beta} \right]^{1 \over n} a^{-\frac{3}{n}}\,+
\frac{4n(2-n)\kappa\eta_{rad}}{3\beta(n-3)^2}
\left[\frac{\eta_{dust}\kappa}{\beta} \right]^{1-n \over n} a^{-\frac{n+3}{n}}
-\frac{4 k n^2}{(n-3)^2}\,a^{-2} \ .
\label{FRV1}
\end{equation}
One should immediately note that this expression, representing
the squared  Hubble parameter, reproduces in the case $n=\beta=1$ -- as
expected -- the standard Friedmann equation. We would like to emphasize also
that (11) becomes singular at $n=3$.\\

It is convenient to rewrite relation (11) is such a way that all
coefficients are dimensionless (density parameters).
Then, the effects of the matter scaling like $a^{-3(1+w)}$, and the radiation
scaling like  $a^{-4}$, can be separated from the effects of the nonlinear
generalization of Einstein gravity ($n \ne 1$):
 
 \begin{eqnarray}
\left(\frac{H}{H_0}\right)^2=
\Omega_{m,0}(1+z)^{3}\frac{2n}{\left(3-n\right)}
\Omega_{nonl,0}(1+z)^{\frac{3\left(1-n\right)}{n}}+ \nonumber \\
+\Omega_{r,0}(1+z)^{4}\frac{4n\left(2-n\right)}{\left(n-3\right)^2}
\Omega_{nonl,0}(1+z)^{\frac{3\left(1-n\right)}{n}} \ .
\label{eq:12a}
\end{eqnarray}
Here:
$\Omega_{m,0}=\frac{\eta_{dust}\kappa}{3 H^2_0}$,
$\Omega_{r,0}=\frac{\eta_{rad}\kappa}{3 H^2_0}$,
$\Omega_{nonl,0}=\left(\frac{\eta_{dust}\kappa}
{\beta}\right)^{\frac{1-n}{n}}$ while $H_0$ denotes the present-day value of
the Hubble function.
%and $\Omega_{k,0}=-\frac{k}{H^2_0 a_0^2}$.
Let us observe  that $\Omega_{nonl,0}$ can be determined also from the 
constraint $H(z=0)=H_{0}$, which easily reduces to:
$$\Omega_{nonl,0}=\left(\frac{2n}{\left(3-n\right)}\Omega_{m,0}+
\frac{4n\left(2-n\right)}{\left(n-3\right)^2}\Omega_{r,0}\right)^{-1} \ .$$

The relation (\ref{eq:12a}) has the form of Friedmann's first integral. 
Therefore, the dynamics of the model can be naturally rewritten in terms 
of a 2D dynamical system of Newtonian type. Its Hamiltonian is:

\begin{equation}
\mathcal{H} \equiv \frac{1}{2}\dot{a}^2+ V(a)=0 \ ,
\label{eq:13}
\end{equation}
while the corresponding equations of motion are:

\begin{align}
\dot{a} &=y \ ,\nonumber \\
\dot{y} &=-\frac{\partial V}{\partial a} \ .
\label{eq:14}
\end{align}
The overdot  differentiats now with respect to the rescaled time variable 
$\tau$, so that $dt= |H_0|d \tau$,  while $V(a)$ is a potential function
for the scale factor $a$ expressed in units of its present value $a_0=1$. 
If $H^2=f(a)$, then the potential function is given by the general formula

\begin{equation}
V(a)= -\frac{1}{2} f(a) a^2 \ .
\label{eq:15}
\end{equation}

For example, the potential function for our model  writes as ($a_0=1$):

\begin{equation}
V(a)=-\frac{1}{2}\left[
\frac{2n}{3-n}\Omega_{m,0}\,a^{-1}
%\left(\Omega_{nonl,0}a^{-\frac{3}{n}\left(1-n\right)}\right)+
+ \frac{4n\left(2-n\right)}{\left(n-3\right)^2}\Omega_{r,0}\,a^{-2}\right]
\Omega_{nonl,0}\,a^{\frac{3}{n}\left(1-n\right)} \ .
%+\Omega_{k,0}\frac{4n^2}{\left(n-3\right)^2
\label{eq:16}
\end{equation}

Recently, in Carloni et al \cite{car}, the cosmological dynamics of $R^n$ 
gravity has been treated in a different phase space with the use of 
qualitative methods for dynamical systems.

The phase portraits for the $\Lambda$CDM model versus our model
with fitted values of $n, \Omega_{m,0}$ parameters (see the next Section)
are illustrated on Figures 1 and 2. Both models are topologically
inequivalent: the phase portrait of $\Lambda$CDM has a structurally
stable saddle critical point, while with nonlinear gravity
one obtains a center. As well-known, the critical point of a center type
is structurally unstable and all trajectories around this point represent
the models, which oscillate  without initial and final singularities.

\begin{figure*}[ht!]
\includegraphics[width=0.9\textwidth]{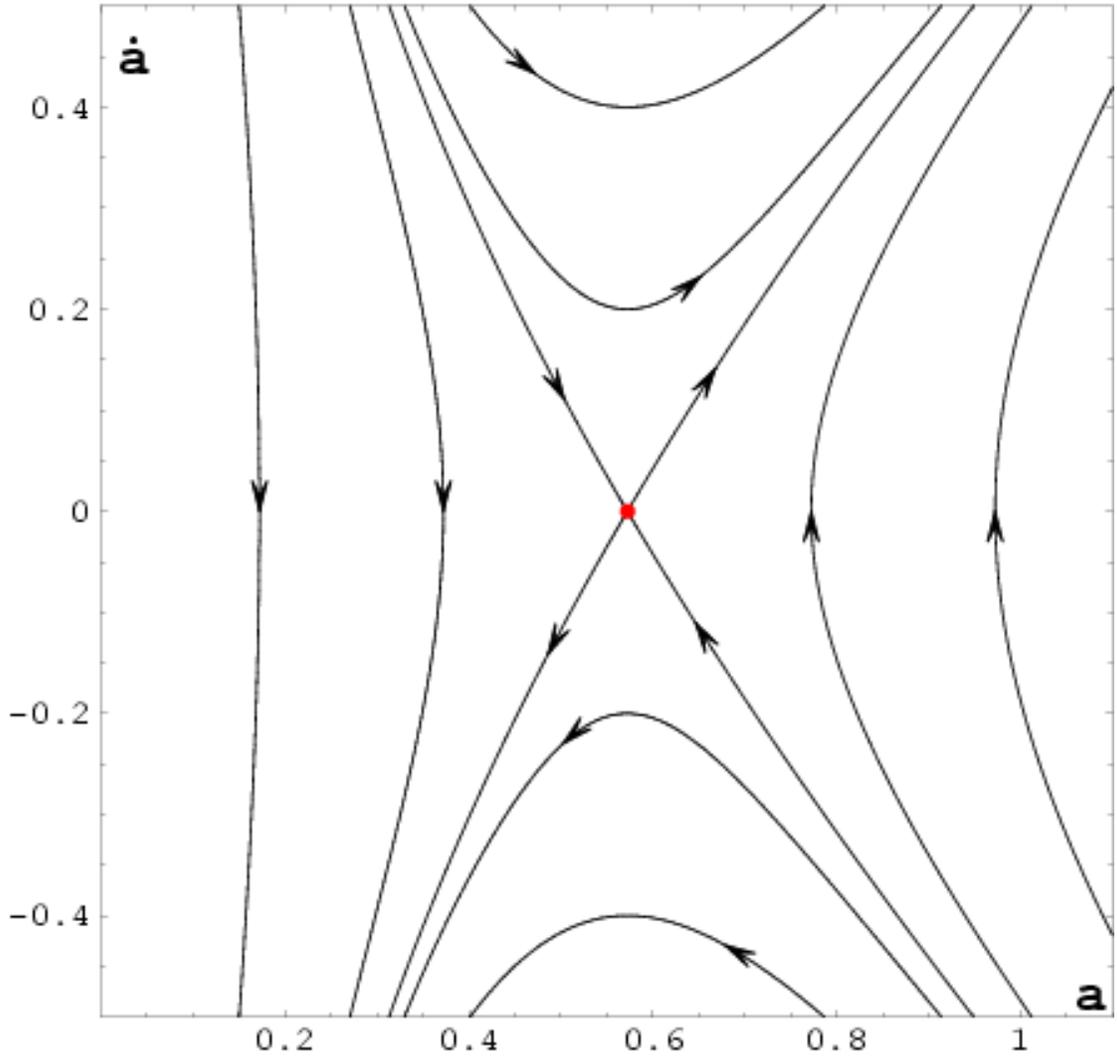}
\caption{The phase portrait for the $\Lambda$CDM model. There is a
single critical saddle-point on the $a$-axis. It represents the
static Einstein universe. The trajectory of flat $k=0$ model divides
all remaining ones into closed (inside) and open (outside) models.}
\label{fig:1}
\end{figure*}

\begin{figure*}[ht!]
\includegraphics[width=0.9\textwidth]{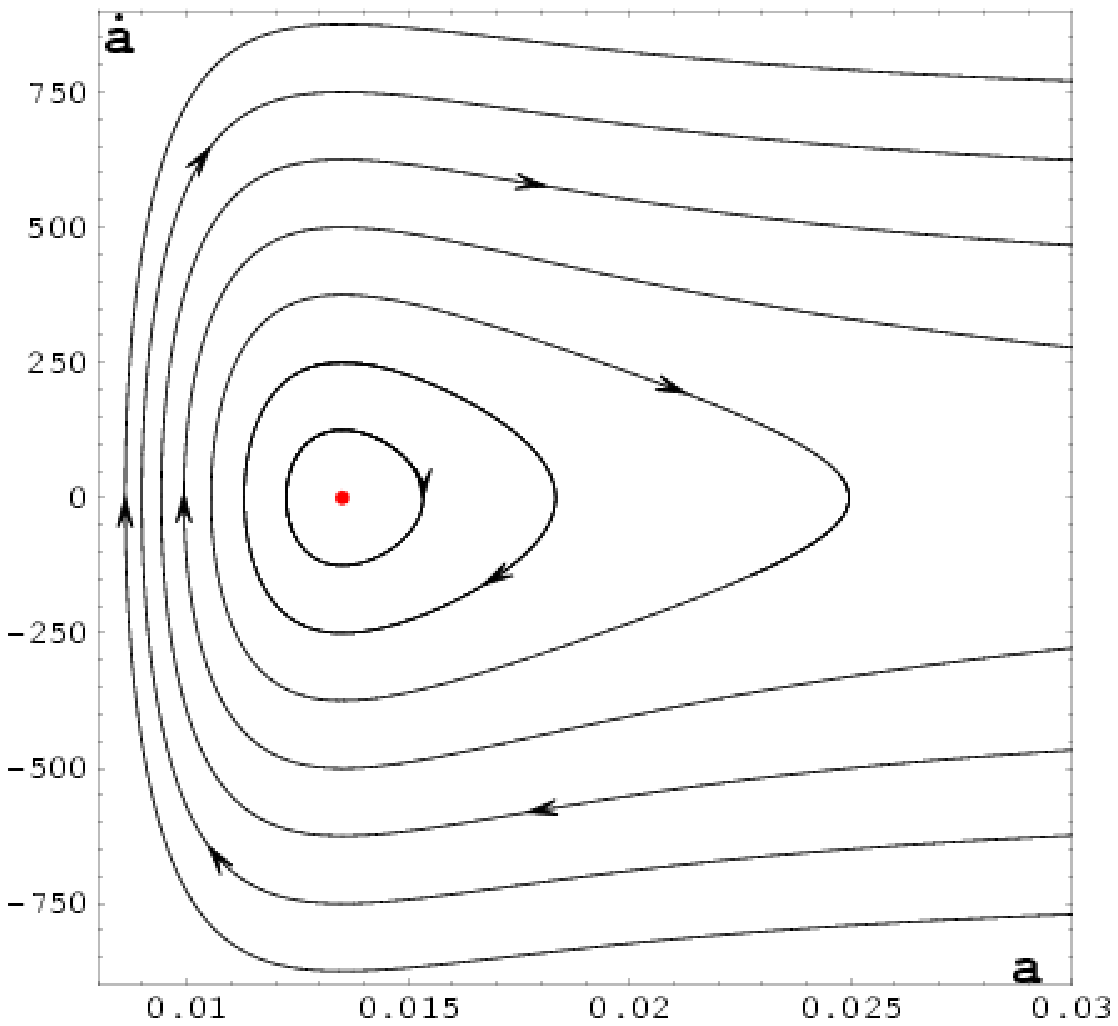}
\caption{The phase portrait for nonlinear gravity with 
${\cal L}
\propto R^n$, $n=2.6$ (from estimation). There is a single critical
point on the $a$-axis - a center;
$a=a_{crit}=\frac{8n(n-2)}{(n-3)^2}
\left(\frac{\Omega_{r,0}}{\Omega_{m,0}}\right)$. The trajectories of
the system lie in the physical region $\{a:a>\frac{a_{crit}}{2}\}$
and represent bouncing evolution. In this scenario, instead of the
big bang singularity of the $\Lambda$CDM model, one has a bounce
$a=a_{min}$, $\ddot{a}>0$. It lies in a neighborhood of the minimum
of the potential function. During the bounce phase, the universe is
still accelerating. Note that if radiation effects vanish, there is
no static critical point on the $a$-axis (formally $a_{crit}=0$ is
allowed for $n>3$).} \label{fig:2}
\end{figure*}

It is interesting that (\ref{eq:12a}) %with $\Omega_{k,0}=0$
after a time reparametrization following the rule: $d\eta=
(1+z)^{\frac{3\left(1-n\right)}{2}}d\tau$ is equivalent to the
standard cosmological model with matter and radiation, with rescaled
values of the corresponding density parameters $\Omega_{m,0}$ and
$\Omega_{r,0}$.

The geometry of the potential function offers the possibility to
investigate the remaining models. On can simply establish some
general relation between the geometry of the potential function and
critical points of the Newtonian systems. In any case, the critical
points lie on the $a$-axis, i.e. they represent the static solution
$y_0=0,\ a=a_0$ so that $\left(\frac{\partial V}{\partial
a}\right)_{a_0}=0$. If $(a_0,0)$ is a strict local maximum of
$V(a)$, it is of the saddle type. If $(a_0,0)$ is a strict local
minimum of the analytical function $V(a)$, it is a centre. If
$(a_0,0)$ is a horizontal inflection point of the $V(a)$, it is a
cusp.

From the fitting procedure we obtain $n > 2$, so the second term in
the potential function is negative (in  contrast to the first term
which is positive). Because the negative radiation term in (\ref{eq:16}) 
can not dominate the first one ($V \le 0$), there is the characteristic 
bounce behavior rather than the initial singularity in the $\Lambda$CDM
model. Moreover, during the bouncing phase the universe is accelerating, 
while for late times it becomes matter dominated and decelerates.
%%%%%%%%%%%%%%%%%%%%%%%%%%%%%%%%%%%%%%%%%%%%%%%%%%%%%%%%%%%%%%%%%%%%%%%%%%%%%%%%%%%%

\section{Distant supernovae as cosmological test}

Type Ia distant supernova surveys suggest that the present Universe
is accelerating \cite{Riess:1998cb,Perlmutter:1998np}. Every year new SNIa
enlarge the available data by more distant objects and lower systematics
errors. Riess et al. \cite{Riess:2004nr} have compiled samples which
become the standard data sets of SNIa. One of them, the restricted ``Gold''
sample of 157 SNIa, is used in our analysis. Recently Astier et al.
\cite{Astier:2005} have compiled a new sample of supernovae, based on
71 high redshifted SNIa discovered during the first year of %the 5-year
Supernovae Legacy Survey. This latest sample of 115 supernovae is used
as our basic sample.

For distant SNIa one can directly observe their apparent magnitude $m$
and redshift $z$. Because the absolute magnitude $M$ %{\bf by\l o  ${\cal M}$ }
is related to the absolute luminosity $L$, the relation
between luminosity distance $d_L$, the observed magnitude $m$ and
the absolute magnitude $M$ has the following form
\begin{equation}
\label{eq:4}
m - M = 5\log_{10}d_{L} + 25 \ .
\end{equation}
It is convenient to use the dimensionless 
parameter $D_L$
\begin{equation}
\label{eq:5}
D_{L}=H_{0}d_{L}
\end{equation}
instead of $d_L$.
Then ~(\ref{eq:4}) can be replaced by
\begin{equation}
\label{eq:6}
\mu \equiv m - M = 5\log_{10}D_{L} + \mathcal{M} \ ,
\end{equation}
where
\begin{equation}
\label{eq:7}
\mathcal{M} = - 5\log_{10}H_{0} + 25 \ .
\end{equation}
We know the absolute magnitude of SNIa from its light curve. The luminosity
distance of supernovae is a given  function of the redshift:
\begin{equation}
\label{eq:8}
d_L(z) =  (1+z) \frac{c}{H_0} \frac{1}{\sqrt{|\Omega_{k,0}|}}
\mathcal{F} \left( H_0 \sqrt{|\Omega_{k,0}|} \int_0^z \frac{d z'}{H(z')} \right) \ ,
\end{equation}
where $\Omega_{k,0} = - \frac{k}{H_0^2}$ and
\begin{align}
%\begin{equation}
\label{eq:9}
\mathcal{F} (x) &=  \sinh (x) \qquad &\text{for} &\qquad k<0 \ ,\nonumber \\
\mathcal{F} (x) &=         x  \qquad &\text{for} &\qquad k=0  \ , \\
\mathcal{F} (x) &=  \sin (x)  \qquad &\text{for} &\qquad k>0 \ .\nonumber
%\end{equation}
\end{align}

Substituting (\ref{eq:8}) back into equations (\ref{eq:4}) and (\ref{eq:6}) 
provides us with an effective tool (Hubble diagram) to test cosmological 
models and to constrain their parameters. Assuming that supernovae
measurements come with uncorrelated Gaussian errors, one can determine the
likelihood function $\mathcal{L}$   from  chi-square
statistic $\mathcal{L}\propto \exp(-\chi^{2}/2)$, where
\begin{equation}
\label{eq:10}
\chi^{2}=\sum_{i}\frac{(\mu_{i}^{\mathrm{theor}}-\mu_{i}^{\mathrm{obs}})^{2}}
{\sigma_{i}^{2}} \ .
\end{equation}
The Probability Density Function (PDF in short) of cosmological parameters
\cite{Riess:1998cb} can be derived from Bayes' theorem. Therefore, one can
estimate model parameters by using a minimization procedure. It is based on
the likelihood function as well as on the best fit method  minimizing $\chi^2$.

For statistical analysis we have restricted the parameter $\Omega_{m,0}$
to the interval $[0,1]$ and $n$ to $[-10.0,10.0]$ (except $n=0$ and 
additionally $n=3$ for $w=0$). Moreover, because of the
singularity at  $n=3, w=0$ (see eq.~(\ref{eq:12a})) we have separated
the cases $n>3$ and $n<3$ for $w=0$ in our analysis. Please note that
$\Omega_{nonl,0}$ is obtained from the constraint $H(z=0)=H_{0}$.

In Figure 3  we present residual plots of redshift-magnitude
relations (Hubble diagram) between the Einstein-de Sitter model
(represented by zero line) and our best fitted  model --- upper
curve  --- and $\Lambda$CDM model --- middle curve. One can observe
systematic deviations between these models at higher redshifts. The
non-linear gravity model predicts that high redshifted supernovae
should be fainter than those predicted by the $\Lambda$CDM model.

\begin{figure*}[ht!]
\includegraphics[width=0.45\textwidth]{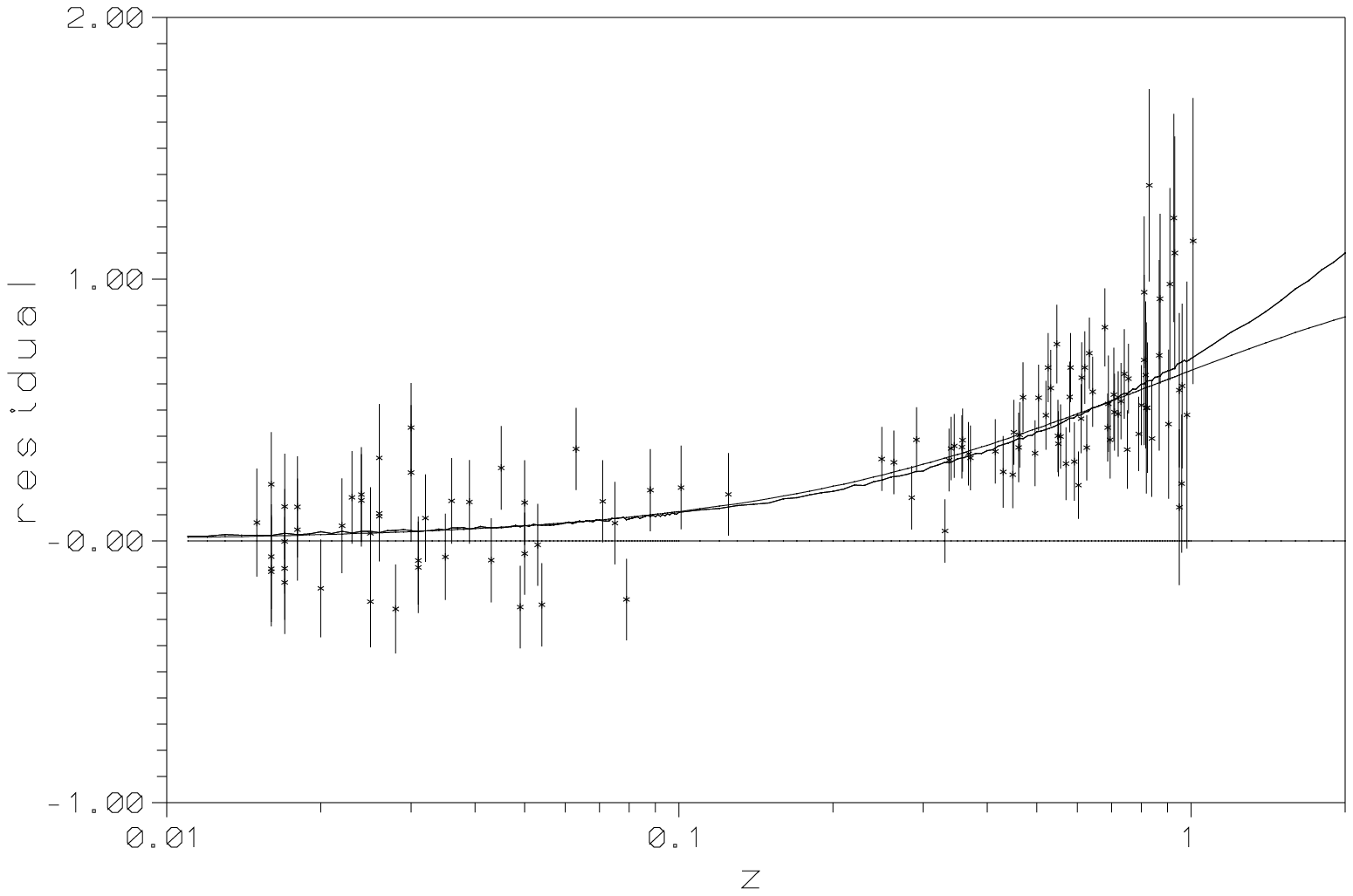}
\includegraphics[width=0.45\textwidth]{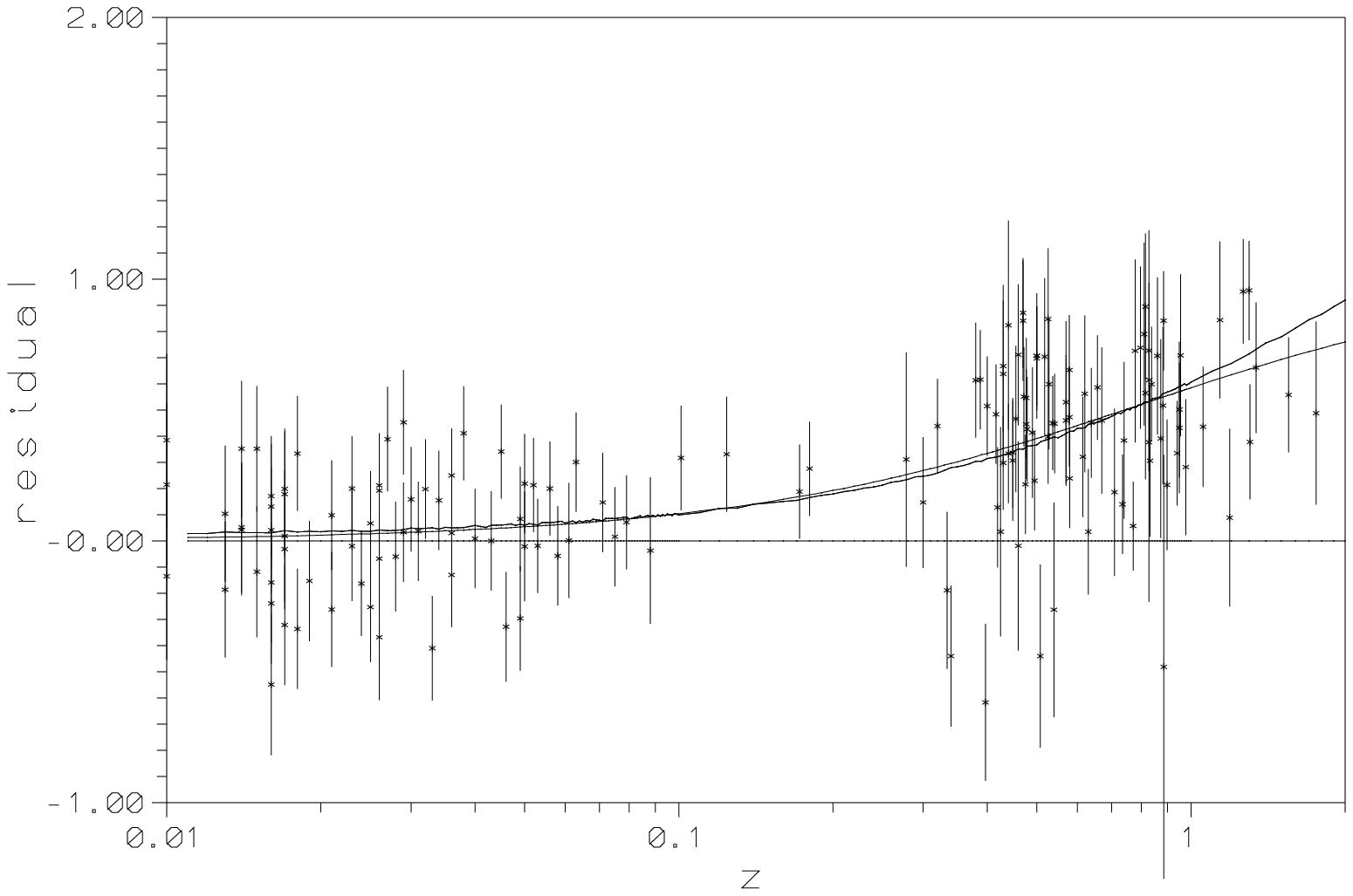}
\caption{Residuals (in mag) between the Einstein-de Sitter model
(zero line), the flat $\Lambda$CDM model (middle curve) and the
non-linear gravity model (upper curve). Results obtained with the
Astier (left panel) and the Riess (right panel) samples.}
\label{fig:3}
\end{figure*}

\begin{figure*}[ht!]
\includegraphics[width=0.45\textwidth]{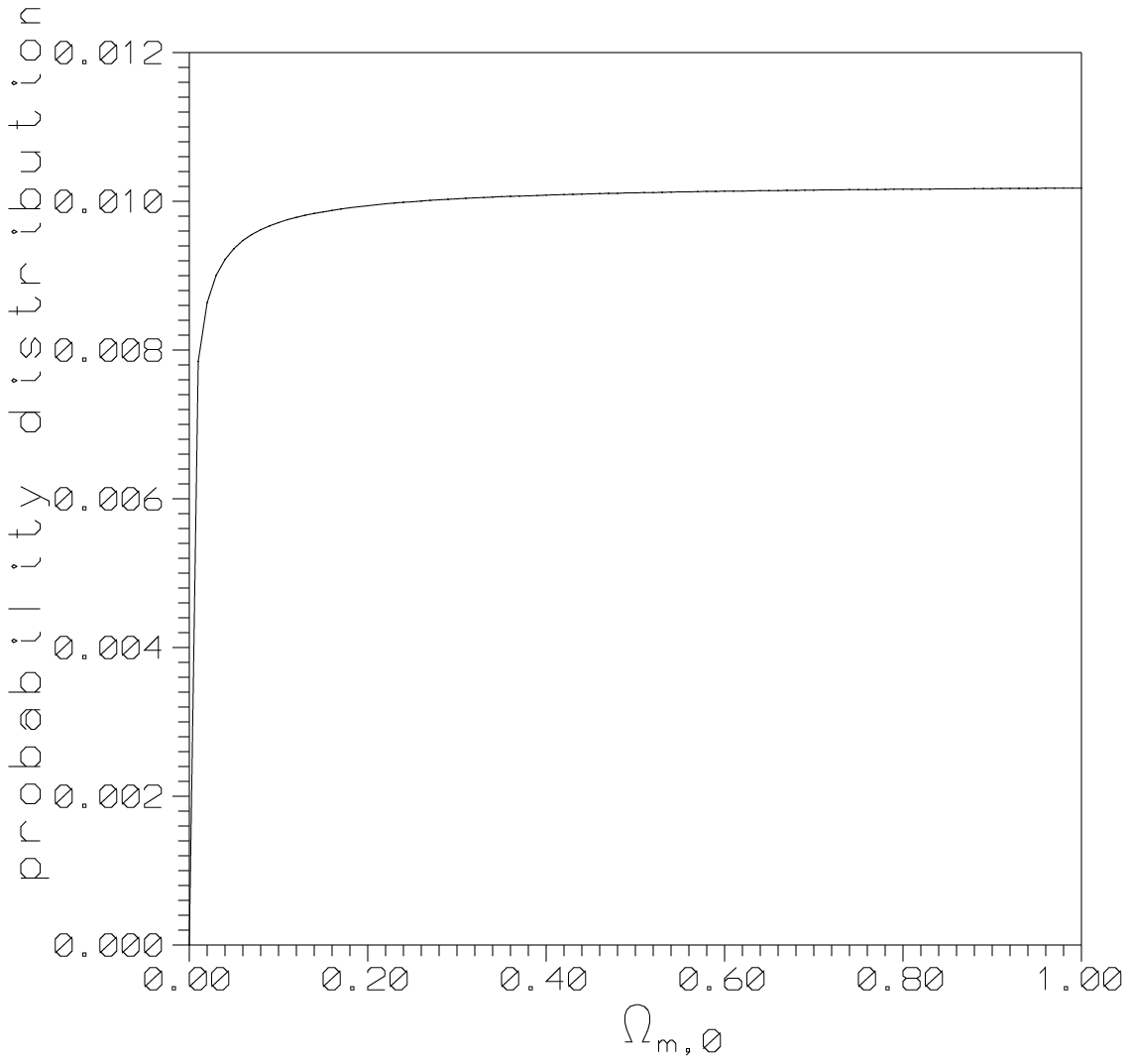}
\includegraphics[width=0.45\textwidth]{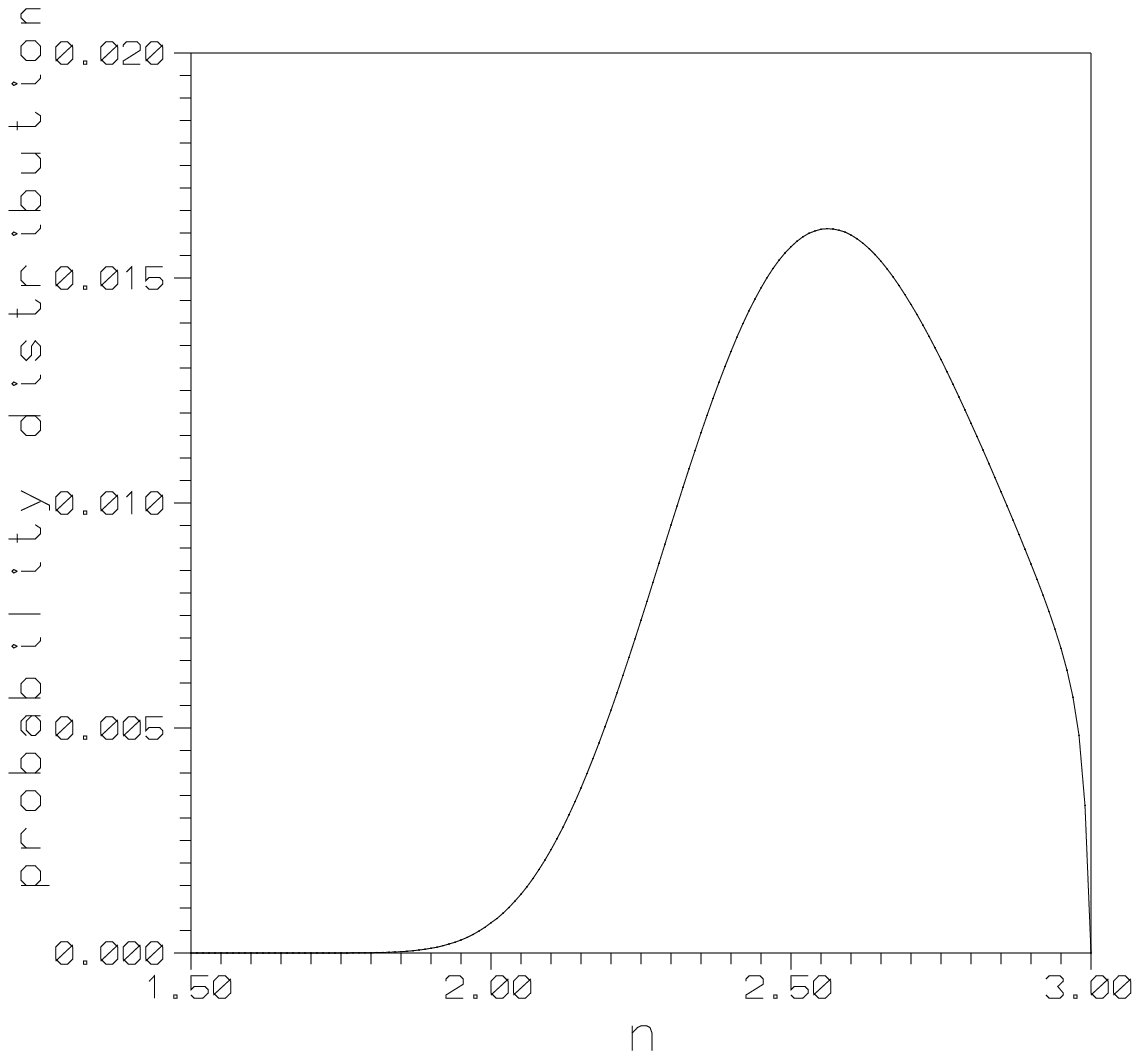}
\caption{PDF obtained  with the Astier sample for the parameters
$\Omega_{m,0}$ and $n$, marginalised over the rest of parameters.
Non-linear gravity model ($w=0$, $n<3$).} \label{fig:4}
\end{figure*}

The results of two fitting procedures performed on Riess and Astier
samples  with different prior assumptions for $n$ are presented in
Tables 1 and 2. In the Table 1 the values of model parameters
obtained from the minimum of $\chi^2$ are given, whereas in Table 2
the results from marginalised probability density functions are
displayed. Please note that we obtained different values of ${\cal M}$
from the Riess versus Astier samples. It is because Astier
et al. assume the absolute magnitude $M=-19.31 \pm 0.03  +
5\log_{10} h_{70}$ \cite{Astier:2005}. For comparison we present
(Table 3) results of statistical analysis for the $\Lambda$CDM 
concordance model.

\begin{table}
\caption{The flat non-linear gravity model with $w=0$. Results of
statistical analysis performed on the Astier versus the Gold Riess
samples of SNIa obtained from $\chi^2$ best-fit. We separately
analysed the case $n>3$ and $n<3$ .} \label{tab:1}
\begin{tabular}{@{}p{1.5cm}rrrrr}
\hline  \hline
sample & $\Omega_{m,0}$ & $\Omega_{nonl,0}$ & $n$ & $\mathcal{M}$ & $\chi^2$ \\
\hline
%\startdata
Gold   &  0.35 &$<0.01$ & 3.001&15.975&180.7     \\
$n<3$  &  0.89 &$ 0.23$ & 2.13 &15.975&181.5     \\
$n>3$  &  0.35 &$<0.01$ & 3.001&15.975&180.7     \\
\hline
Astier &  0.01 &$-1.47$ & 3.11 &15.785&108.7     \\
$n<3$  &  0.98 &$ 0.08$ & 2.59 &15.785&108.9     \\
$n>3$  &  0.01 &$-1.47$ & 3.11 &15.785&108.7     \\
\hline
%\enddata
\end{tabular}
\end{table}

\begin{table}
\caption{The flat non-linear gravity cosmological model ($w=0$). The
values of the parameters obtained from marginal PDFs calculated on
the Astier versus the Gold Riess samples. Because of the singularity
at $n=3$ we separately analyze the cases $n>3$ and $n<3$ .}
\label{tab:2}
\begin{tabular}{@{}p{1.5cm}rrrr}
\hline \hline
sample & $\Omega_{m,0}$ & $\Omega_{nonl,0}$ & $n$ & $\mathcal{M}$ \\
\hline
Gold  &$ 0.01^{}$ & $0.26^{}_{}$ & $2.11^{}_{}$ & $15.955^{+0.03}_{-0.03}$\\
$n<3$ &$ 1.00_{}$ & $0.26^{}_{}$ & $2.11^{}_{}$ & $15.955^{+0.03}_{-0.03}$\\
$n>3$ &$ 0.01^{}$ & $-0.01^{}_{}$& $3.001^{}_{}$& $15.955^{+0.03}_{-0.03}$\\
\hline
Astier&$ 0.01^{}$ & $0.09^{}_{}$ & $2.56^{}_{}$ & $15.785^{+0.03}_{-0.03}$\\
$n<3$ &$ 1.00^{}$ & $0.09^{}_{}$ & $2.56^{}_{}$ & $15.785^{+0.03}_{-0.03}$\\
$n>3$ &$ 0.01^{}$ &-$0.01^{}_{}$ & $3.01^{}_{}$ & $15.785^{+0.03}_{-0.03}$\\
\end{tabular}
\end{table}

\begin{table}
\caption{Results of statistical analysis of the $\Lambda$CDM flat
model performed on the Astier versus the Gold Riess samples of SNIa
as a minimum $\chi^2$ best-fit.} \label{tab:3}
\begin{tabular}{@{}p{1.5cm}rrrr}
\hline
sample & $\Omega_{m,0}$ & $\Omega_{\Lambda,0}$ & $\mathcal{M}$ & $\chi^2$ \\
\hline \hline
%\startdata
Gold   &  0.31 & 0.69 & 15.955& 175.9     \\
Astier &  0.26 & 0.74 & 15.775& 107.8     \\
\hline
%\enddata
\end{tabular}
\end{table}

The best fit (minimum $\chi^2$) gives $n \simeq 2.6$ with the Astier et al. 
sample versus $n \simeq 2.1$ with the Gold sample. In Figure 4 we present 
PDF obtained  with the Astier sample for the parameters $\Omega_{m,0}$ and 
$n$ for non-linear gravity model, (case $n<3$ marginalised over the rest of 
parameters). Please note that from Fig. 4 we
obtain a very weak dependence of PDF on the matter density parameter if only
$\Omega_{m,0} \ge 0.05$.

In Figure 5, confidence levels on the  plane $(\Omega_{m,0},n)$, for
non-linear gravity model, for the case $n<3$ marginalized over ${\cal M}$
are presented.

Recently Eisenstein et al.  have analyzed baryon oscillation peaks
detected in the  Sloan Digital Sky Survey (SDSS) Luminosity Red Galaxies 
\cite{Eisenstein:2005}. They found

\begin{equation}
\label{eq:116}
A \equiv =\frac{\sqrt{\Omega_{m,0}}}{E(z_1)^{\frac{1}{3}}}
\left(\frac{1}{z_1\sqrt{|\Omega_{k,0}|}}
\mathcal{F} \left( \sqrt{|\Omega_{k,0}|} \int_0^{z_1} \frac{d z}{E(z)} \right)
\right)^{\frac{2}{3}} \ ,
\end{equation}
so that $E(z) \equiv H(z)/H_0$ and $z_1=0.35$  yield $A=0.469 \pm 0.017$.
The quoted uncertainty corresponds to one standard deviation, where a Gaussian
probability distribution has been assumed.
These constraints could also be used for fitting cosmological parameters
\cite{Astier:2005,Fairbairn:2005}. We obtain from this test the values of the
model parameters $\Omega_{m,0}=0.28$, $\Omega_{nonl,0}=0.33$ and $n=2.53$ for
a best fit. In Figure 6 we show the region allowed by the baryon
oscillation test on the plane $(\Omega_{m,0},n)$ for non-linear gravity model
(for the case $n<3$). In Figure 7 we present combined confidence levels, 
obtained from the analysis \cite{Fairbairn:2005} of both data sets. We find 
that the model favours $\Omega_{m,0} \simeq 0.3$ and $n \simeq 2.6$.

\begin{figure*}[ht!]
\includegraphics[width=0.9\textwidth]{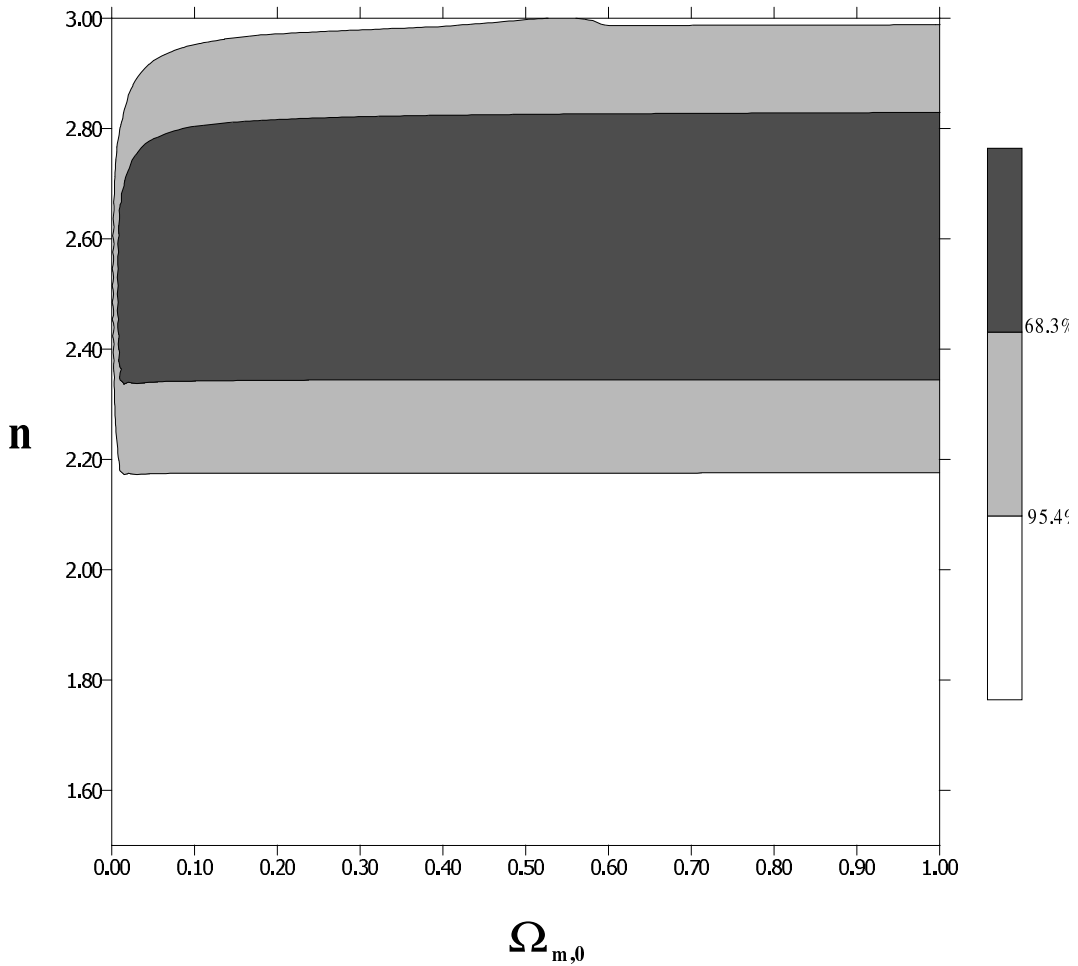}
\caption{The flat non-linear gravity model ($w=0, n<3$). Confidence
levels on the $(\Omega_{m,0},n)$ plane, marginalised over ${\cal
M}$, obtained from SNIa Astier sample.} \label{fig:5}
\end{figure*}

\begin{figure*}[ht!]
\includegraphics[width=0.9\textwidth]{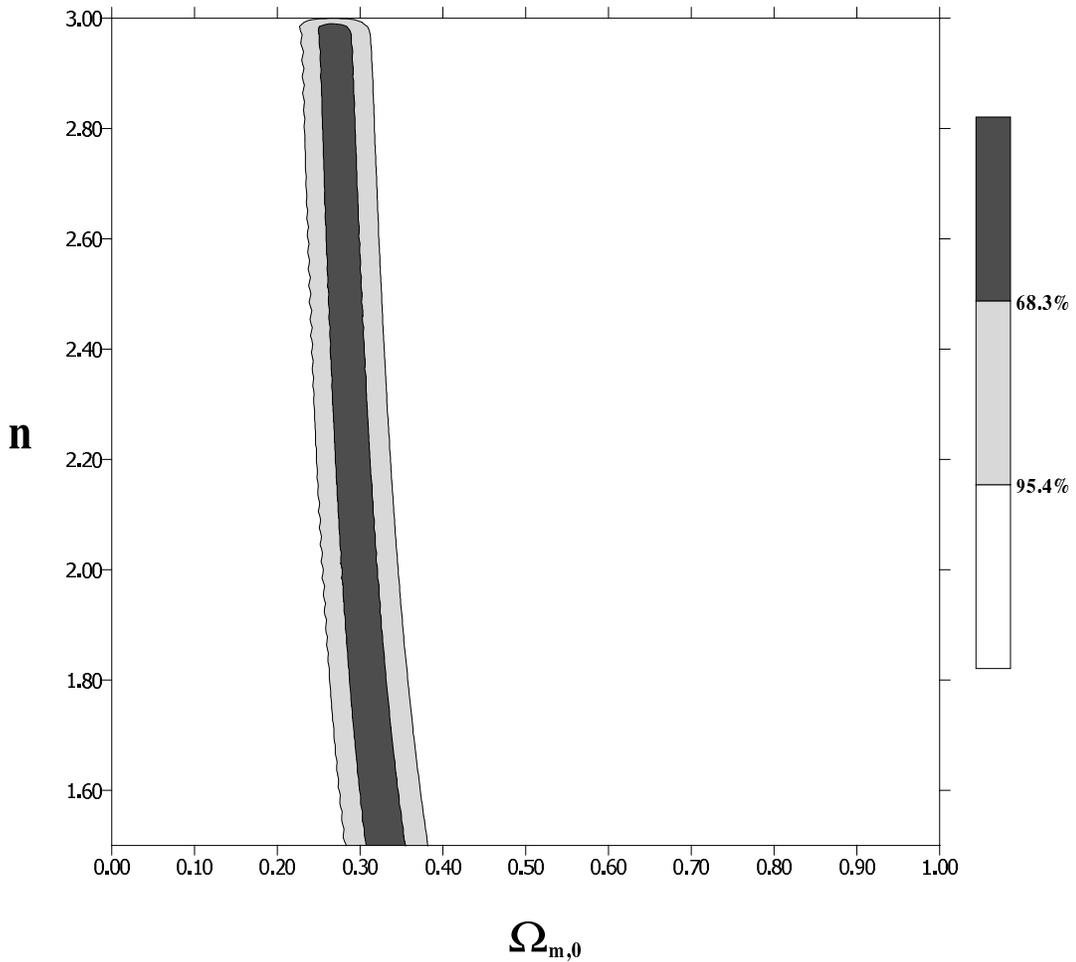}
\caption{The flat non-linear gravity model ($w=0, n<3$). Confidence
levels on the $(\Omega_{m,0},n)$ plane obtained from baryon
oscillation peaks.} \label{fig:6}
\end{figure*}

\begin{figure*}[ht!]
\includegraphics[width=0.9\textwidth]{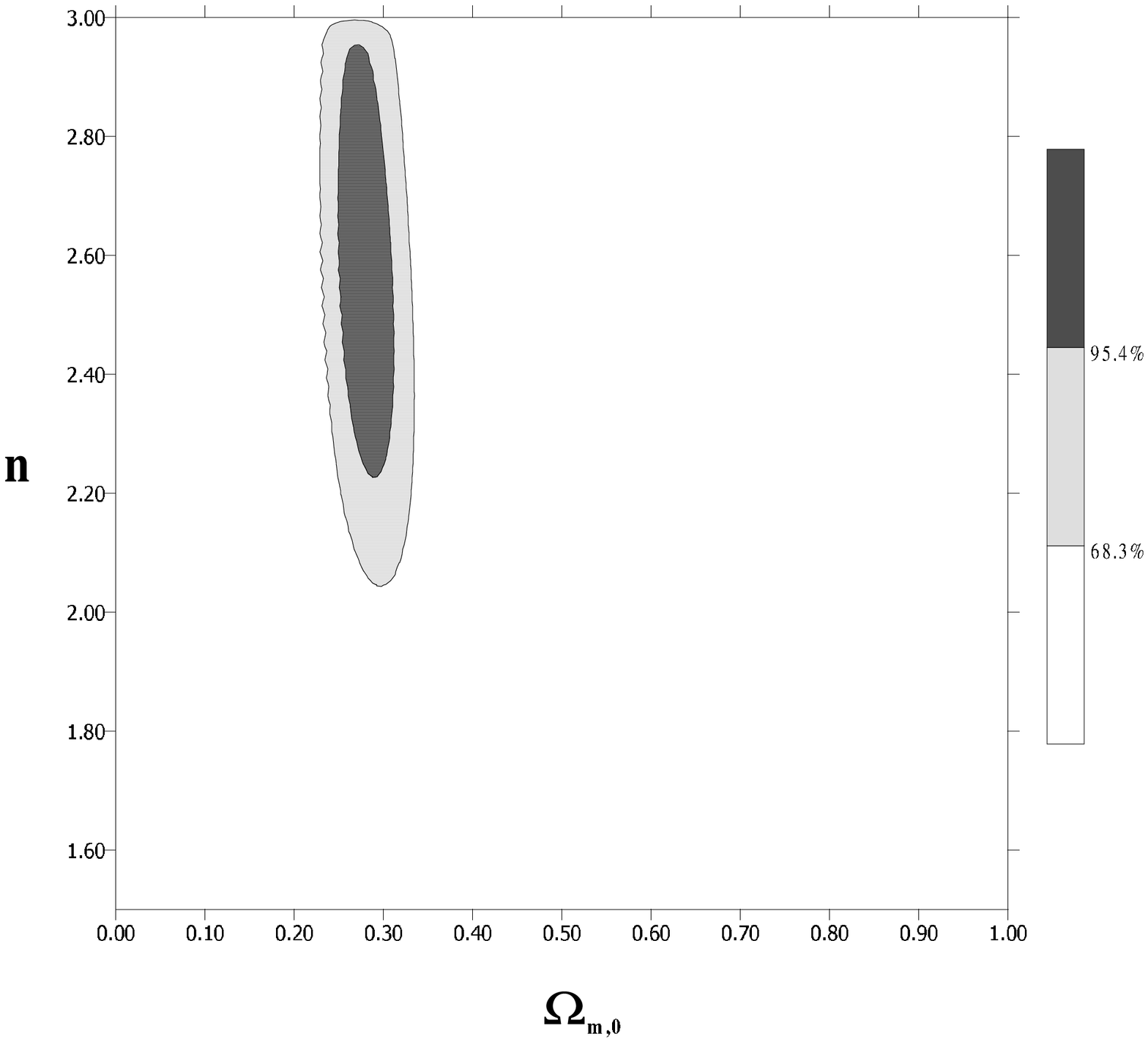}
\caption{The flat non-linear gravity model ($w=0, n<3$). Common
confidence levels on the plane $(\Omega_{m,0},n)$ obtained from SNIa
Astier sample and baryon oscillations.} \label{fig:7}
\end{figure*}

In modern observational cosmology, one encounters
the so-called degeneracy problem: many models with dramatically
different scenarios (big bang or bounce, big-rip or de Sitter phase)
agree with the present day observational data. Information
criteria for model selection \cite{Liddle:2004nh} can be used,
in some subclass of dark energy models, in order to overcome this degeneracy
\cite{Godlowski05,Szydlowski06}. Among these, Akaike
(AIC) \cite{Akaike:1974} and  Bayesian (BIC) information
criteria  \cite{Schwarz:1978} are the most popular. From these
criteria one can determine several essential model
parameters, providing the preferred fit to the data \cite{Liddle:2004nh}.

The AIC \cite{Akaike:1974} is defined by
 \begin{equation}
\label{eq:111}
\mathrm{AIC} = - 2\ln{\mathcal{L}} + 2d \ ,
\end{equation}
where $\mathcal{L}$ is the maximum likelihood and $d$ the number of
model parameters. The best model, with a parameter set providing the preferred
fit to the data, is that which minimizes the AIC.

The BIC introduced by Schwarz \cite{Schwarz:1978} is defined as
\begin{equation}
\label{eq:112}
\mathrm{BIC} = - 2\ln{\mathcal{L}} + d\ln{N} \ ,
\end{equation}
where $N$ is the number of data points used in the fit. While AIC tends to
favor models with a large number of parameters, the BIC
penalizes them more strongly, so the later provides a useful approximation
to the full evidence in the case of no prior on the set of model parameters 
\cite{Parkinson:2005}.

The effectiveness of using these criteria in the current cosmological
applications has been recently demonstrated by Liddle \cite{Liddle:2004nh}.
Analyzing CMBR WMAP satellite data \cite{Bennett:2003bz}, he found the number 
of essential cosmological parameters to be five. Moreover, he came to important
conclusion that spatially-flat models are statistically preferred to close
models as it was indicated by the CMBR WMAP analysis (their best-fit value is
$\Omega_{tot,0} \equiv \Sigma_i \Omega_{i,0} = 1.02 \pm 0.02$ at
$1\sigma$ level).

In the paper by Parkinson et~al. \cite{Parkinson:2005}, also the usefulness of
Bayesian model selection criteria in the context of testing for double
inflation with WMAP was demonstrated. These criteria were also used
recently by us to show that models with the big-bang scenario are rather
preferred over the bouncing scenario \cite{Szydlowski:2005qb}.

Please note that both information criteria make no absolute sense and
only the relative values between different models are physically interesting.
For the BIC a difference of $2$ is treated as a positive evidence
($6$ as a strong evidence) against the model with larger value of BIC
\cite{Jeffreys:1961,Mukherjee:1998wp}.
Therefore one can order all models, which belong to the ensemble of dark
energy models, following the AIC and BIC values. If we do not find any
positive evidence from information criteria, the models are treated as
identical, while eventually additional parameters are treated as not 
significant. Therefore, the information criteria offer a possibility to 
introduce a relation of weak ordering among considered models.

\begin{table}
\caption{Results of AIC and BIC  performed on
the Astier versus the Gold Riess samples of SNIa.}
\label{tab:4}
\begin{tabular}{@{}p{3.5cm}rr}
\hline
sample &  AIC & BIC \\
\hline \hline
%\startdata
$\Lambda$CDM  Gold     & 179.9 & 186.0 \\
$\Lambda$CDM  Astier   & 111.8 & 117.3 \\
Non-Lin.Grav. Gold     & 186.6 & 195.8 \\
Non-Lin.Grav. Astier   & 114.7 & 122.9 \\
\hline
%\enddata
\end{tabular}
\end{table}

For comparizing the $\Lambda$CDM and the non-linear gravity models the
results of  AIC and BIC are presented in Tables
\ref{tab:4}. Note  that for both samples we obtain  with
AIC and BIC for the $\Lambda$CDM model smaller values than for
non-linear gravity. We use a Bayesian framework to compare the
cosmological models, because they automatically penalize models with
more parameters to fit the data. Based on these simple information
criteria, we find that the SNIa data still favour the $\Lambda$CDM
model, because under a similar quality of the fit for both models,
the $\Lambda$CDM  contains fewer parameters.

It is interesting that both models give different predictions for the 
brightness of the distant supernovae (see Fig. 3). The model of 
modified non-linear gravity predicts that very high redshift supernovae 
should be fainter than predicted  by $\Lambda$CDM. So, we can expect 
future SNIa data to allow us discriminating finally between these two models.

\section{Conclusion}

The main subject of our paper has been to confront the simplest class of
non-linear gravity models versus the observation of distant type Ia supernovae 
and the recent detection of the baryon acoustic peak in the Sloan Digital 
Sky Survey data. We find strong constraints on two independent model parameters
($\Omega_{m,0},\, n$). If we assume n=1, then we obtain the standard Einstein
de Sitter model filled by both matter and radiation. We estimate model 
parameters using standard minimization procedure based on the likelihood 
function as well as the best fit method. For deeper statistical analysis, 
we have used AIC and BIC information criteria of model comparison and 
selection. Our general conclusion is that non-linear gravity fits well 
(both SNIa and baryon oscillation data).
In particular we conclude:
\begin{enumerate}
\item Analysis of SNIa Astier data shows that values of $\chi^2$ statistic are
comparable for both $\Lambda$CDM and best fitted non-linear gravity model.
\item The non-linear gravity models with $n<2$ can be excluded by combined
analysis of both SNIa data  and baryon oscillation peak detected in the
SDSS Luminous Red Galaxy survey of Eisenstein at al. \cite{Eisenstein:2005}
on $2 \sigma$ confidence level.
\item From SNIa data we obtain a weak dependence of the quality of fits
on the value of density parameter for matter ($\Omega_{m,0}$).
However, the combined analysis allowed only value of $\Omega_{m,0}$ well tuned
to its canonical value $\Omega_{m,0}=0.3$. This value of course is in
good agreement with present extra-galactic data \cite{Peebles:2002gy}.
\item We use the Akaike and Bayesian information criteria for comparison and
discrimination between the analyzed models. We find these criteria still
to favour the $\Lambda$CDM model over non-linear gravity, because (under the
similar quality of the fit for both models) the $\Lambda$CDM model contains
one parameter less.
\item The Hubble diagram implies that very high redshifted supernovae
($z \ge 1.5$) should be fainter in non-linear gravity model than those
predicted by $\Lambda$CDM. So, future SNIa data can allow us finally
to discriminate between these two models.
\item The standard general relativity models with $n=1$ (without cosmological 
constant) can be excluded by SNIa data on $17 \sigma$ level (as the E-deS 
model).
\item The non-linear cosmology can therefore be treated as a
serious alternative versus cosmology with dark energy of unknown nature.
\end{enumerate}

\section{Acknowledgements}
Authors thank Dr A.G. Riess and Dr P. Astier for the detailed explanation of 
their supernovae samples. We also thank Dr M. Fairbairn, Dr S. Capozziello
and Dr K. Just for helpful discussion. M. Szydlowski acknowledges the support 
by KBN grant no. 1 P03D 003 26. A. Borowiec is supported by KBN grant 
no. 1 P03B 01 828.

\end{document}